\documentclass[10pt,journal,compsoc]{IEEEtran}
\usepackage{hyperref}
\usepackage{amsmath}
\usepackage{amssymb}
\usepackage[utf8]{inputenc}
\usepackage[english]{babel}
\usepackage{cleveref,multirow}
\usepackage{xspace}
\usepackage{subfloat}
\usepackage{subcaption}
\usepackage{algorithm}
\usepackage{graphicx} %
\usepackage[noend]{algpseudocode}
\usepackage{booktabs}
\usepackage{tabularx}
\usepackage[symbol]{footmisc}

\newcommand{\Biocode}{B\textsc{io}C\textsc{ode}\xspace}
\newcommand{\Load}{L\textsc{oad}\xspace}
\newcommand{\Save}{S\textsc{ave}\xspace}
\newcommand{\Swap}{S\textsc{wap}\xspace}
\newcommand{\Clear}{C\textsc{lear}\xspace}
\newcommand{\Set}{S\textsc{et}}
\newcommand{\Rewind}{R\textsc{ewind}}

 \newcommand{\Createedge}{G\textsc{enerate\_edge}\xspace}
 \newcommand{\Randomnode}{R\textsc{andom\_node}\xspace}
 \newcommand{\Randomedge}{R\textsc{andom\_edge}\xspace}
 \newcommand{\Newnode}{N\textsc{ew\_node}\xspace}
\newcommand{\Influenceneighbors}{I\textsc{nfluence\_neighbours}}
\newcommand{\Detachfrominfluenced}{D\textsc{isconnect\_from\_influenced}\xspace}
\newcommand{\Attachtoinfluenced}{C\textsc{onnect\_to\_influenced}\xspace}
\newcommand{\Clearinfluenced}{C\textsc{lear\_influenced}\xspace}
\newcommand{\Skipinstruction}{S\textsc{kip\_instruction}}

\newcommand{\bse}{\begin{subequations}}
\newcommand{\ese}{\end{subequations}}
\newcommand{\ben}{\begin{enumerate}}
\newcommand{\een}{\end{enumerate}}
\newcommand{\bens}{\begin{enumerate*}}
\newcommand{\eens}{\end{enumerate*}}
\newcommand{\be}{\begin{equation}}
\newcommand{\ee}{\end{equation}}
\newcommand{\bea}{\begin{eqnarray}}
\newcommand{\eea}{\end{eqnarray}}
\newcommand{\baa}{\begin{eqnarray*}}
\newcommand{\eaa}{\end{eqnarray*}}
\newcommand{\bc}{\begin{center}}
\newcommand{\ec}{\end{center}}

\newcommand{\argmax}{\operatornamewithlimits{argmax}}
\newcommand{\avg}{\operatornamewithlimits{avg}}

\newcommand{\norm}[1]{\left\lVert#1\right\rVert}

\ifCLASSOPTIONcompsoc
  \usepackage[nocompress]{cite}
\else
  \usepackage{cite}
\fi
\ifCLASSINFOpdf
\else
\fi

\begin{document}
\title{\Biocode: A Data-Driven Procedure to Learn the Growth of Biological Networks}


\author{Emre~Sefer
  %~\IEEEmembership{Member,~IEEE,}
\IEEEcompsocitemizethanks{\IEEEcompsocthanksitem Emre Sefer is with the Department
  of Computer Science, Ozyegin University, Istanbul, Turkey.\protect\\

 %He was also with JP Morgan Applied AI Research, New York City, USA. \protect \\

E-mail: emre.sefer@ozyegin.edu.tr
%\IEEEcompsocthanksitem J. Doe and J. Doe are with Anonymous University.}% <-this % stops an unwanted space
% \thanks{Manuscript received April 19, 2005; revised August 26, 2015.}
}}

\IEEEtitleabstractindextext{%
\begin{abstract}

Probabilistic biological network growth models have been utilized for many tasks including but not limited to capturing mechanism and
dynamics of biological growth actitivies, null model representation, capturing anomalies, etc. Well-known examples of these probabilistic models are Kronecker model,
preferential attachment model, and duplication-based model. However, we should frequently keep developing new models to better fit and
explain the observed network features while new networks are being observed. Additionally, it is difficult to develop a growth model each time we
study a new network. In this paper, we propose \Biocode, a framework to automatically discover novel
biological growth models matching user-specified graph attributes in directed and undirected biological graphs. \Biocode designs basic set of instructions which are common enough to
model a number of well-known biological graph growth models. We combine such instruction-wise representation with a genetic algorithm
based optimization procedure to encode models for various biological
networks. We mainly evaluate the performance of \Biocode in discovering models for
biological collaboration networks, gene regulatory networks, metabolic
networks, and protein interaction networks which features such as assortativity, clustering coefficient, degree distribution closely
match with the true ones in the corresponding real biological networks. As shown by the tests on the simulated graphs, the variance of the
distributions of biological networks generated by \Biocode is similar
to the known models variance for these biological network types.
  
\end{abstract}

}

\maketitle

\IEEEdisplaynontitleabstractindextext

% \fi
\IEEEpeerreviewmaketitle

\IEEEraisesectionheading{\section{Introduction}\label{sec:introduction}}

Research of the dynamics by which temporal evolution of biological
networks occur is a key component in understanding how such biological
networks operate. Especially, understanding the dynamics and appearance of topological
features in biological networks such as modularity, assortativity,
disassortativity, and shrinking diameter is notably important. Creating idealized graph growth models is a successful method in
understanding how such graph features emerge in the first
place. Examples of such idealized graph growth models are preferential attachment models
[e.g.~\cite{Barabasi1999, Rak2020}], duplication/mutation
models~\cite{Bhan2002, Ispolatov2005, Sole2002, Teichmann2004,
  Vazquez2011, Jasra2015}, the Kronecker model~\cite{Leskovec2010,
  Seshadhri2013}, forest fire model~\cite{Leskovec2005}, and other models~\cite {Piva2021, Falkenberg2020, Callaway2001,
  Dorogovtsev2000, Kim2008, Kumar2006, Przulj2009, Huang2017}. Those models describe the biological networks growth mechanistically
and probabilistically. In common, those models express such growth by
union of various operations such as node duplication, node expansion, node/edge creation, node/edge deletion, and influence propagation.

Besides simulating realistic biological network growth, such graph
growth models are used for other purposes in different applications where growth dynamics do not
need to be interpretable in some applications. As an example, growth models may help in inferring the historical networks~\cite{Navlakha2011}, may be helpful in
anonymization~\cite{Leskovec2010}, can be utilized to test the
performance of lengthy large-scale graph methods, may be used as null models to detect anomalous graph features.

The first theoretical studies on network models has begun with
Erdos-Rényi model~\cite{Erdos1960}. Following research on network
models found small-world~\cite{Watts1998} and a scale-free node degree
distribution~\cite{Barabasi1999} properties as frequent real world
network attributes and designed growth models to generate such
properties. Subsequent growth models included additional properties as objectives
across different domains. One such property is clustering
coefficient for protein interaction networks~\cite{Vazquez2011,
  Teichmann2004, Voordeckers2015} which resulted in DMC~(duplication,
mutation, with complementarity) model. Another property is shrinking
diameter for an temporally evolving social network~\cite{Leskovec2005} which ended up in forest fire model. Following attempts~\cite{Akoglu2009, Palla2010} have designed manual
models which fit multiple extra properties simultaneously and have
generated real-word like graphs. More recent models try to also match richer node features in addition
to real world network topologies~\cite{Sun2009, Kim2012}. It is a challenging task to create a feasible, parsimonious, network
growth model that fits well to the data. As we study large-scale and different types of networks, we will
identify new properties which require developing novel growth models. Nevertheless, the degree custom-made growth models model the desired
network properties will depend on model designer's capabilities and creativity.

In this paper, we come up with a formal characterization of network
growth models which encode well-known and frequently-used network
growth models in addition to many more undiscovered
models. Additionally, we introduce an optimization framework which can automatically
discover novel models that better fit the desired network attributes
in the aforementioned formal setting. Models learned by the proposed framework can generate many sample
networks across different classes matching input properties. These
learned models are relatively easy to understand and interpretable
with an effort of certain degree. In many scenarios, graph motifs can be frequently mined in the set of generated growth models as generated models are in general distinct
and better fit the desired properties. These mined motifs are in
general successfull in modeling a specific network attribute.
%Burayi cevirmedim!!!
%Additionally, the ease with which good-fitting models can be found can be used as a measure of
%the ubiquity of that feature among graph growth mechanisms.
Lastly, in numerous situations, computationally derived growth
models outperform human-designed models in matching real-world attributes.

Among the existing work, only a few research has focused on
automatically designing network growth models. Some of the earlier frameworks can adapt existing models to novel
graph data via recalculating model parameters governing graph
dynamics. As an example, Kronecker graph model parameters can be estimated better by integrating Markov Chain Monte Carlo~(MCMC) method
to its parameter estimation, especially for matching several very
large network attributes~\cite{Leskovec2010}. Another example is
estimating the parameters for additional recursive growth
models~\cite{Akoglu2009}. However, those methods are restricted to estimating network model
parameters and they cannot mimick novel network growth dynamics.~\cite{Middendorf2005} focus on selecting the best models among a few
current models. According to this work, DMC model fits protein-protein
interaction graphs the best~\cite{Vazquez2011}. Nonetheless, their approach does not fit parameters for current models
and does not generate novel models.

Here, we design a framework called \Biocode to address those
inadequacies via encoding fundamental graph operations and other graph
model defining structures as instructions operating in a virtual
machine with multiple registers. Intuitively, providing an effective
set of atomic instructions and network growth structures are our main
motivation. A series of such consecutive instructions define a network growth
dynamics iteration, and recurrent iterations of these series of instructions
temporally grow a graph. One of our main contribution in \Biocode is that only a number of
operations are enough to model a duplication model, a forest-fire-like model, preferential attachment model, and supposedly
more growth models. Among these operations, $4$ of them have parameters whereas the
rest of operations are parameterless. Moreover, the machine operated by \Biocode operations has only $3$
registers. Such smaller machine design restrains the total number of
candidate programs which allows for an efficient search of the solution
space by a genetic algorithm.

\Biocode allows us to learn biological network growth models automatically and quickly which assures a number of biological
networks key features. \Biocode models often outperform human-designed models in fitting to
the fundamental topological graph features of clustering coefficient,
assortativity, and degree distribution. Particularly, model learned by \Biocode on yeast protein interaction
networks~\cite{Janjic2014} generate graphs better than the popular DMC
model which simulates these protein interaction networks in terms of
agreeing to the observed the degree distribution and clustering
coefficient values. Additionally, we can outperform Kronecker model with the best
parameters~\cite{Leskovec2005} in terms of generating graphs that match the degree distribution and assortativity of a recent
biological co-authorship network~\cite{Bulik-Sullivan2012}. Lastly, the models identified by \Biocode is better than a Kronecker
model in terms of simultaneously matching node degree distribution,
assortativity, and clustering coefficient of a gene regulatory
network. In our settings, the graphs generated by \Biocode learned models are more diverse than
the ones generated by the competing human-designed models, showing
that models generated by \Biocode are correctly random network models.

Even though the proposed \Biocode framework generates unattributed
graphs, the process suggested by \Biocode is quite common and
widespread. We can extend \Biocode to different graph classes. The technique pointed by \Biocode allows for automatic and more
systematic graph growth dynamics study.

\section{The \Biocode Framework}

We come up with \Biocode framework where we can express growth models
programmatically and briefly. A register machine together with $15$ instructions executing
on the register machine are defined. Every series of machine instructions is a correct program concisely encoding a
biological graph growth model. Basic and particular operations impacting the topological graph features are included
in \Biocode instruction set. Few instructions are included to direct
the program flow and manage registers. Mainly, instructions that are natural growth model structures are included in \Biocode instruction set. The selected instructions may
represent a number of unknown and existing biological models.

\Biocode machine changes an evolving graph's topology while it
executes a program. Every single execution of \Biocode instructions in a program outlines
a growth process single step. We execute \Biocode program from scratch till the end $t$ times in order to
evolve a network for $t$ time steps. $t$ is linked to the output graph size, and $t$ is an implicit
parameter for each \Biocode program. \Biocode machine registers are
filled randomly with the graph nodes between the successive program
calls, modeling the successive growth steps. When combined with a number of randomized instructions, this
randomization between successive program runs aids \Biocode to encode
probabilistic growth models as \Biocode programs. Consequently, the same
program's separate executions almost always generate dissimilar biological networks.

%\subsection{A register machine}
\subsection{\Biocode registers}

\Biocode executes instructions on a $3$ register virtual
machine. These registers are r$0$, r$1$, r$2$ which can store positive
integers. Register values mainly correspond to node IDs, even though their
values may also correspond to parameters used by several instructions. Register may take a special value NIL showing the
register is idle. \Biocode keeps a program counter, PC which displays
the presently running instruction. Once an instruction is run, program counter is increased in order to
maintain a sequential execution of the program as long as one of the
control flow instructions updates the program counter. As achieved by \Rewind instruction below, \Biocode programs can modify
themselves to support looping to a certain extenrt. As in a
traditional computer program, program is terminated when program
counter location passes beyond the program length.

Let $V$ be the evolving graph's nodes, \Biocode incorporates a limited
amount of memory $L : V \rightarrow V$ which is able to store a single
node ID for each node in $V$. Here, $L(v)$ value on vertex $v$ may not necessarily be $v$'s node ID,
instead it may be an another node's ID. This ability of graph vertices to store IDs of any other graph vertex
is the key factor on \Biocode programs spreading a vertex's influence
in the evolving graph~(See Section~\ref{sec:instructionset} for
details). $v$ does not have a label when $L(v) = $NIL. It is possible to have more complicated influence operations and
memory models. However, our experiments show that good agreement can
be achieved across multiple different settings by our proposed minimal design.

\subsection{\Biocode instruction set}\label{sec:instructionset}

%\subsubsection{Design of Instruction set design}

It is a difficult and long-established problem to design instructions on virtual and physical processors. We carefully included an operation in \Biocode instruction set if such
operation represents a fundamental graph operation. Each \Biocode instruction is easily comprehensible and resemble the operations seen
in human-created graph growth models. Union of those instructions may end up in growth models which can generate graphs with the required features.
%Cevirmedin
%Among multiple possible instruction sets, the union of \Biocode operations is just one example.
%The combination of instructions used here is but one example among many possible instruction sets
\Biocode instructions may be expanded by including further
instructions to incorporate novel graph growth processes. Optimizing a hard objective becomes relatively easier via a carefully-designed good instruction set.
However, totally resolving instruction set design problem is not this
paper's focus. Instead, via our experiments, we show that a single instruction
set in Table~\ref{tab:main} performs quite accurately for many
biological graph classes.

\begin{table*}[ht] 
\caption{Complete set of \Biocode instructions}
\begin{center}
\begin{tabular}{lll}
\toprule
Operation Type & \parbox[b]{3.0cm}{\centering Operation} &\parbox[b]{4cm}{\centering Definition} \\
  \midrule
  \multirow{5}{*}{Register}
  &\Clear r$2$ & set r$2$ to NIL \\ 
  &\Swap & Swap r$0$ and r$1$ contents \\
&\Save & Clone register r$0$ content to r$2$ \\
 &\Load & Clone register r$2$ content to r$0$ \\
  &\Set$(i)$ & Clone vertex ID to r$2$ \\ \hline
  \multirow{2}{*}{Control flow}
  &\Skipinstruction$(p)$ & Pass over the following operation \\ 
  &\Rewind$(r, i)$ & Go back $r$ lines $i$ times \\ \hline            
  \multirow{4}{*}{Influence}
  &\Clearinfluenced & Clean all tags in $L$ \\ 
 &\Detachfrominfluenced & Delete edges to the neighbours tagged with $u$ \\ 
&\Attachtoinfluenced & Add edges to neighbours tagged with $u$ \\
&\Influenceneighbors$(p)$ & Tag neighbours with $u$ \\ \hline
  \multirow{4}{*}{Graph} 
  & \Createedge & Generates an edge \\
& \Newnode & Introduces a new vertex \\
  &\Randomedge & Randomly selects an edge \\ 
&\Randomnode & Randomly selects a vertex \\
\bottomrule
\end{tabular}
\end{center}
\label{tab:main}
\end{table*}

% \begin{table*}[ht] 
% \caption{Complete set of \Biocode instructions}
% \begin{center}
% \begin{tabular}{lll}
% \toprule
% Operation Type & \parbox[b]{3.0cm}{\centering Operation} &\parbox[b]{4cm}{\centering Definition} \\
%   \midrule
%  \multirow{5}{*}{Register} &\Set$(i)$ & Clone vertex ID to r$2$ \\
% &\Save & Clone register r$0$ content to r$2$ \\
% &\Load & Clone register r$2$ content to r$0$ \\
% &\Swap & Swap r$0$ and r$1$ contents \\
% &\Clear r$2$ & set r$2$ to NIL \\ \hline
%  \multirow{2}{*}{Control flow}  &\Rewind$(r, i)$ & Go back $r$ lines $i$ times \\
%                &\Skipinstruction$(p)$ & Pass over the following operation \\ \hline
% \multirow{4}{*}{Influence}  &\Influenceneighbors$(p)$ & Tag neighbours with $u$ \\
% &\Attachtoinfluenced & Add edges to neighbours tagged with $u$ \\
% &\Detachfrominfluenced & Delete edges to the neighbours tagged with $u$ \\ 
% &\Clearinfluenced & Clean all labels in $L$ \\ \hline
%   \multirow{4}{*}{Graph} & \Newnode & Introduces a new vertex \\
% & \Createedge & Generates an edge \\
% &\Randomnode & Randomly selects a vertex \\
% &\Randomedge & Randomly selects an edge \\ 
% \bottomrule
% \end{tabular}
% \end{center}
% \label{tab:main}
% \end{table*}

\subsubsection{\Biocode instructions}

We can divide \Biocode instructions in Table~\ref{tab:main} into four
groupings: $1$- Register instructions, $2$- Control flow instructions,
$3$- Graph instructions, and $4$- Influence instructions. Among these groupings, the first two groupings focus on managing
\Biocode state and managing the \Biocode program's control flow as
suggested by their names. The third and fourth groupings focus on
transforming the evolving graph's topology. See Table~\ref{tab:main}
below for a full operations list. Next, we characterize the impact of these operations on the
resulting generated graphs and on \Biocode machine state. Section~\ref{sec:3} discusses the expressibility of a number of graph growth models by those operations.

\textbf{Register instructions:} One can directly manage the contents of $3$
registers by the register instructions. The \Clear r$2$ adjusts the r$2$'s value to NIL. The \Swap
instructions swaps r$0$ and r$1$'s values. The \Save instruction puts r$0$'s value also into r$2$. Contrarily,
\Load copies r$2$'s value into r$0$. The \Set$(i)$ operation puts integer $i$ to r$2$.

\textbf{Control flow instructions:} \Biocode operations order of
execution is modified by these control flow instructions. \Skipinstruction$(p)$ moves the program counter by $2$ with
probability $p$. In this case, the following instruction is
conditionally run with probability $1 - p$ by such probabilistic
movement. Another instruction in this category, \Rewind$(r, i)$, models for
loop-like behaviours. Its first critical argument $r$ specifies how
manu times program counter must be decreased when \Rewind$(r, i)$ is
run. In another words, this parameter also models how far program counter must go in
reverse direction. Its second argument $i$ defines how many times the operations must be
run. Whenever \Biocode executes \Rewind$(r, i)$, it is updated via
decreasing $i$ by $1$. Whenever $i$ becomes $0$, program counter value
will not be rewound as \Biocode will stop execution. These \Rewind$(r, i)$ parameters are reset within successive program runs.

\textbf{Graph instructions:} The graph instructions primarily modify topology of
graph. The \Createedge instruction generates an edge in the graph. \Biocode
retrieves the node IDs from r$0(u)$ and r$1(v)$ registers, and
generates $\{u, v\}$ edge. \Createedge instruction does not alter the register states, so this
instruction does not have an impact if $\{u, v\}$ edge is already part
of the graph. The \Newnode instruction generates a new vertex in the evolving
graph. \Biocode retrieves vertex ID from register r$0$. The \Randomedge instruction uniformly and randomly chooses an edge
$\{u, v\}$ in the graph, and puts corresponding vertices $u$ into r$0$
and $v$ into r$1$. Finally, \Randomnode instruction picks a vertex randomly and uniformly and puts this vertex to r$0$.

\textbf{Influence instructions:} Graph vertices can have an influence
upon other nodes by influence instructions. $L(v) = u$ means vertex
$u$ influences$v$. Such influence mechanism is essential in generating
graphs with different types of features. One example is homophily in which common topological neighborhoods are shared by vertices to a certain degree. A vertex can influence a subset of vertices in its neighbourhood by the main influence instruction, \Influenceneighbors$(p)$. While running the \Influenceneighbors$(p)$ instruction, \Biocode retrieves the vertex ID $u$ from r$0$ where $u$ turns into the influential or central vertex. Afterwards, \Biocode propagates this $u$ mark to $u$'s every neighbour $v$
by assigning $L(v) = u$ probabilistically. Such probabilistic assignment takes place independently for each such
vertex with probability $p$. In turn, every newly marked node $v$ marks its neighbours having content
$u$ with probability $p^{d(u, v)}$, where $d(u, v)$ is the distance of
shortest path between $u$ and $v$. In this case, vertices $v$ such that $d(u, v) < $r$2$ might be
impacted by this influence instruction unless r$2$ is NIL. When r$2 = $NIL, the influence instruction keeps executing till the probabilistic
process terminates and so process does not mark any more vertices.

\begin{figure}[h!]
\begin{center}
\includegraphics[width=8.0cm]{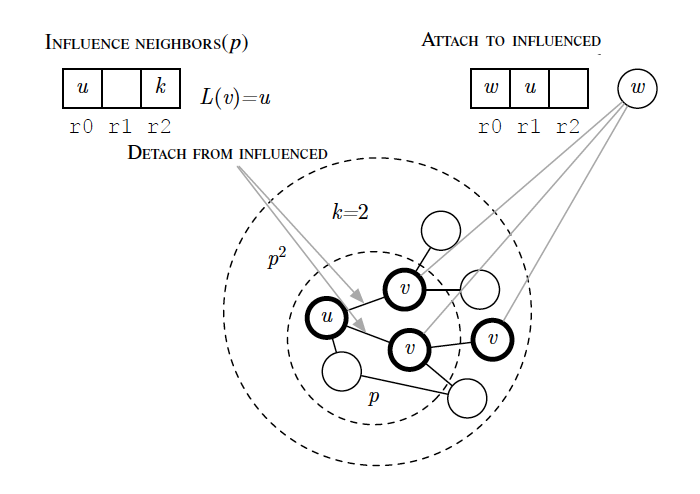}
\caption{The summary of $3$ influence instructions. First of all,
  vertex $u$ puts an influence mark on its neighbours with probability $p$. Then, the influenced
  neighbours $v$ propagates the influence to their neighbours with
  probability $p^{2}$. When vertex $u$ separates from its marked neighbours, $2$ gray edges
shown by the gray arrows will be deleted from the graph. Lastly, $w$ may connect to other vertices $u$ has put an
influence upon.
}
\label{fig:1}
\end{center}
\end{figure}

There are $3$ further instructions that \Influenceneighbors$(p)$ instruction
operates together with: \Clearinfluenced instruction removes $L$ values such that the following
instructions may operate with a clean memory. The \Detachfrominfluenced instruction retrieves a vertex $u$ from r$0$
register, and deletes all edges $\{u, v\}$ satisfying $L(v) = u$. \Attachtoinfluenced generates edges between the vertex in r$0$, $w$, and all vertices marked with
the content of r$1 = u$. \Attachtoinfluenced generates edges $\{w, v\}$ for all $v$ such
that $L(v) = u$. Making the two vertices neighbourhoods more like each
other is a common mechanism provided by \Attachtoinfluenced. Figure~\ref{fig:1} represents all those influence instructions.

\section{Representing Existing Models}\label{sec:3}

We show the general applicability of \Biocode by showing its
expressive power on $3$ well-studied biological network growth models:
forest fire~(FF)~\cite{Leskovec2005}, duplication and mutation
with complementarity~(DMC)~\cite{Vazquez2011},
Barabási-Albert~(B-A)~\cite{Barabasi1999}. We code \Biocode programs
matching these models key features. Those biological growth models illustrate different topological
aspects by matching different styles of realistic biological
networks. As an example, graphs generated by the DMC model exhibit a wide spectrum of
clustering coefficients matching the ones seen in protein-protein
interaction networks. Similarly, graphs generated by the FF model show
densification power law attribute and shrinking diameter while they
evolve. Although there are major variations in these growth models dynamics
and in the graphs features they generate, rather elementary \Biocode programs
can represent those models by using the same set of basic operations. \Biocode operations are reused over separate models indicating their
high-quality in disclosing a variety of network growth dynamics.

\subsection{Barabási-Albert}

According to B-A growth process, newy added nodes attach to higher-degree nodes
with a higher probability.~\cite{Barabasi1999}. B-A process produces
networks with scale-free distribution which is frequently observed in
real-world biological and social networks. According to the scale-free
distribution, too many low-degree vertices in the distribution are
followed by few very high degree vertices.

Algorithm~\ref{alg:1} defines a \Biocode program that similarly mimicks the B-A
model. Even though fine differences between the original B-A model
and the program exist, the graphs generated by B-A model resemble the
ones produced by \Biocode program. The B-A model's fundamental part is
defined in lines $3$–$5$. Among these lines, the \Randomedge operation in line $3$ selects an
edge that has high-degree vertices at its endpoints with high
probability. The probability of a randomly selected edge $e$ containing vertex $u$
is relative to $u$'s degree: $\frac{d(u)}{E} = 2\frac{d(u)}{\sum_{v
    \in v}d(v)}$. While $e$ is chosen, operations in lines $4$ and $5$ randomly select an
endpoint for $e$, ensuring vertex selection within $e$ is without
bias. The model defined by Algorithm~\ref{alg:1} picks up vertices relative
to their degree as required by B-A model with a small difference;
contrary to B-A model, as \Biocode program runs, vertex degrees $d(u)$
are modified. Then, instruction in line $7$ connects the newly introduced vertex to
$u$ accomplishing the newly introduced vertex's preferential
attachment. The following \Rewind\xspace instruction loops over this process
such that the newly introduced vertex is connected to $i$ current vertices.

\begin{algorithm}
\caption{B-A}
\begin{algorithmic}[1]
    \State \Newnode \Comment{Generate a new vertex $u$}
    \State \Save
    \State \Randomedge\Comment{Randomly pick up an edge $e$}
    \State \Skipinstruction$(0.5)$\Comment{Randomly select vertex $v$ of $e$}
    \State \Swap
    \State \Load
    \State \Createedge \Comment{Generate an edge between vertices $u$ and $v$}
    \State \Rewind$(5, i)$\Comment{Randomly attach newly introduced
      vertex to $i$ current vertices}
  \end{algorithmic}
  \label{alg:1}
\end{algorithm}

\subsection{Duplication and Divergence}

DMC model~(Duplication and mutation with complementarity
model)~\cite{Vazquez2011} focuses on generating graphs that mimick
protein interaction graphs topological attributes. Network evolves by
the duplication of current vertices in DMC model. The DMC model has $q_{con}$ and $q_{mod}$ parameters controlling the
network growth as follows: Every newly introduced, duplicated vertex $u$ selects an anchor vertex $v$ and
connects to all neighbours of $v$. For every vertex $w$ adjacent to both $v$ and $u$, an edge is randomly
selected attaching $w$ either to $v$ or $u$, and selected edge is
deleted with $q_{mod}$ probability. Lastly, $u$ and $v$ are connected with $q_{con}$ probability by
introducing an edge between them. The frequent occurence of gene duplication is the main motivation behind
such network growth dynamics especially in protein interaction
networks, where genes synthesizing proteins within the genome are
frequently duplicated. At the beginning, the duplicated genes are identical copies
so the resulting proteins keep all of the interactions seen in the
original protein. Nonetheless, once duplication is over, the interactions between the original and duplicated
genes begin to differentiate as the evolutionary pressure on genes in keeping the original interactions is
decreased. We design Algorithm~\ref{alg:2} in \Biocode which approximates the DMC
model quite closely.

Algorithm~\ref{alg:2} introduces the DMC model coded in \Biocode that
is somewhat different than the one proposed by Vazquez et
al.~\cite{Vazquez2011}. In our process, we cannot exactly simulate the
process of choosing the common neighbours of $v$ and $u$ with $q_{con}$
probability, and removing the edge to either of them. Though, we can accomplish such dynamics similarly by influencing the common neighbours
of each vertex with probability $\frac{q_{mod}}{2}$ as shown in lines
$8$ and $10$ once $v$’s neighborhood are duplicated to $u$. In this case, influence operation behaves precisely same as the
traditional DMC operation only if the influenced neighbours do not
intersect with each other. A neighbouring vertex might be influenced
by both $v$ and $u$. Complementarity attribute of DMC model means that the edge to either
$v$ or $u$ is deleted, but not both. In our corresponding \Biocode program, complementarity is kept as the
program will overwrite the mark on the common vertex, assuring program
can only delete one of the edges $\{v, w\}$ and $\{u, w\}$. That
process may end up with $q_{mod}$ values which results in a marginally
different impact in \Biocode procedures. However, \Biocode produced DMC graphs exhibit clustering
coefficients~(section~\ref{sec:5.3}) and Zipf plots similar to the
original DMC, that are the key attributes DMC model creators have
stressed out in their paper. Besides, the graphs generated by \Biocode Algorithm~\ref{alg:2}
exhibit clustering coefficients and Zipf plots similar to the ones
found in yeast protein interaction graph. So, even though there are fine differences, the \Biocode
algorithm~\ref{alg:2} keeps the fundamental components and features
of the true DMC model.

\begin{algorithm}
\caption{DMC}
\begin{algorithmic}[1]
\State \Randomnode \Comment{Place a randomly selected vertex $v$ in $r0$}
\State \Set$(1)$ \Comment{Set r$2$ ($k$-hop for influence) to $1$}
\State \Influenceneighbors$(1.0)$ \Comment{Influence neighhbours of $v$}
\State \Swap \Comment{Swap r$0$ and r$1$}
\State \Newnode \Comment{Introduce a vertex $u$ to the graph and place it in r$0$}
\State \Attachtoinfluenced \Comment{Attach/Connect vertex $u$ to influenced vertices}
\State \Clearinfluenced
\State \Influenceneighbors$(\frac{q_{mod}}{2})$ \Comment{Influence the
  neighbours of $u$}
\State \Swap
\State \Influenceneighbors$(\frac{q_{mod}}{2})$ \Comment{Influence the
  neighbors of $v$}
\State \Detachfrominfluenced \Comment{Remove edges from vertex $v$}
\State \Swap
\State \Detachfrominfluenced \Comment{Remove edges from vertex $u$}
\State \Clearinfluenced
\State \Skipinstruction$(1.0 - q_{con})$ \Comment{Pass over addition
  of edge $\{u, v\}$}
\State \Createedge \Comment{Create the edge with $q_{con}$ probability}
\end{algorithmic}
\label{alg:2}
\end{algorithm}

\subsection{Forest Fire}

\cite{Leskovec2005} introduced the forest fire (FF) model to better model
the frequently observed real-world network properties such as temporal
densification of the graph under a certain parameter range, shrinking
diameter, and in and out-degree scale-free degree distributions. The forest fire model can be easily and intuitively explained from
graph growth prospect. Here, we introduce a hardly changed model which
applies to undirected graphs. Once a newly introduced vertex $u$ is added to the graph, such vertex
selects a current vertex $v$ randomly and uniformly which then acts as
an agent and the edge between $v$ and $u$ is joined. Afterwards, forest fire model draws a natural number $n$ from a
geometric distribution with success probability $b$, and $v$'s $n$
neighbours are selected and burned. FF model introduces an edge from vertex $u$ to each of those burned vertices,
and the procedure of choosing a number of neighbouring vertices and
burning these vertices is recursively rerun.

FF model is encoded by \Biocode program in Algorithm~\ref{alg:3}. Graphs generated by the \Biocode program and the FF model are same
in terms of fundamental graph features. Particularly, the graphs generated by Algorithm~\ref{alg:3} display densification
power law and shrinking diameter for certain range of parameters during graph evolution over time.

\begin{algorithm}
\caption{FF}
\begin{algorithmic}[1]
\State \Randomnode  \Comment{Place a random vertex in r$0$}
\State \Clear r$2$ \Comment{Clean r$2$ contents for complete graph influence}
\State \Influenceneighbors$(b)$ \Comment{Propagate influence recursively as breadth-first}
\State \Swap \Comment{Put the random vertex into r$1$}
\State \Newnode \Comment{Introduce newly created vertex, $u$}
\State \Createedge
\State \Attachtoinfluenced \Comment{Connect/Attach vertex $u$ to influenced vertices}
\end{algorithmic}
\label{alg:3}
\end{algorithm}

\section{Learning \Biocode Models}

By expressing biological network growth models in terms of a number of
\Biocode operations, learning a biological graph growth model over \Biocode can be
expressed formally as an optimization problem over the \Biocode
instructions search domain. \Biocode uses genetic programing methods to learn a set of
instructions generating biological networks that mimick given set of
graph attributes as close as possible. \Biocode encodes those network
attributes within an individual \Biocode program's fitness function. Recovering the formerly introduced growth models is not the main goal of
\Biocode learning process, instead we focus on learning programs which
grow graphs that represent specific graph classes as measured by
particular similarity metrics.

\subsection{Constructing a fitness function}

\Biocode defines an attribute collection $x = [x_1, x_2, \ldots, x_m]$ as
a $m$-long feature vector where each entry $x_i$ can represent a
single scalar value such as assortativity, or it can
represent a vector of values such as multiple independent samples of the graph's
effective radius during its evolution. Representing the fundamental and necessary graph attributes \Biocode
will match as part of growth model is the main objective of attribute
collection step. Let $s_{l}(.,.)$ be a user-defined similarity metric between the collections
$l^{th}$ attributes. To calculate the similarity between any two attribute collections of the
same dimension, we define a possibly weighted metric $s(x^i; x^j)$ as in:
\begin{equation}\label{eq:1}
s(x^i; x^j) = \sum_{t=1}^{m}\,w_{l}s_{l}(x_{l}^i, x_{l}^j)
\end{equation}
where similarity measure $s_{l}(.,.)$ can simply be inverse of the difference between two
attributes for single scalar values. Or, it can also represent a metric of the distribution similarities for nonscalar attributes. There are two conditions on $s(x^i; x^j)$: 1- $s(x^i; x^j)$ must get
the maximum value when $x^j = x^i$, 2- $s(x^i; x^j)$ must be a monotonically
non-decreasing function of the similarity between the two attributes. One can weight every attribute separately by the weights $w_{l}$ in
Eq.~\ref{eq:1} which then causes optimization process to prefer some
attributes more than the others. We use $w_{l} = 1$ for all $l$ in our
experiments.

The fitness of a \Biocode program is defined by using Eq.~\ref{eq:1}. Let $x^{P}$ be a random variable defining the attribute collection for
the graph produced by non-deterministic program $P$, and let $x^{T}$
be a target attribute collection. Then, our problem becomes searching for optimal $P^{*}$ such that:
\begin{equation}\label{eq:2}
P^{*} = \argmax_{P}\,\mathbb{E}[s(x^P, x^T)],
\end{equation}
where the expectation is calculated over $P$'s multiple
non-deterministic executions. In this case, \Biocode searches for the optimal program $P^{*}$ such
that the attributes of the graph produced by $P^{*}$ should be the
most similar to the attributes of the graph given by $x^{T}$ according
to similarity measure $s(.,.)$. This optimization problem cannot be easily tackled since number of
candidate programs in the search space is massive. \Biocode handles this problem efficiently by using a genetic
programming algorithm that is proven to be quite useful across difficult
optimization problems.

\subsection{Optimization with genetic algorithms}\label{sec:4.2}.

In \Biocode, we apply the optimization procedures in genetic algorithm
by using the ECJ package~\cite{Scott2019}. We utilize ECJ's capabilities for following reasons: 1- Parallel
evaluation of individuals inside a generation, 2- Customization of the
breeding and selection processes for more than one subpopulations, 3-
Applying NSGA-II multi-objective optimization~\cite{Deb2002}, and
4- Handling various representations for variable and fixed length
genomes. Every candidate individual in the genetic program describes a
program. We calculate the fitness of all individual candidates in the constant-size
population at each generation. \Biocode evaluates the fitness of each program by executing the
program for $k$ iterations, and then compares the program's attribute
vector $x^{P}$ with the target attribute vector. That evaluation process is rerun for $M$ times, and mean of the
calculated results is presented such that the program $P$'s fitness is:
\begin{equation}\label{eq:3}
F(P) = \avg\,s(x^{P}, x^{T})
\end{equation}
As an alternative, we may calculate the mean for each
$s_{l}(x^{i}_{l}; x_{l}^j)$ in Eq.~\ref{eq:1} as an independent objective,
and apply a multi-objective optimization procedure such as NSGA-II~\cite{Deb2002}.

As part of \Biocode optimization process, we breed individual programs and the programs blend with each other by a two-point crossover operation.
This crossover operation vary the programs length and content. As part of each generation's final step, individual programs contest
in a tournament where two randomly selected programs are compared consecutively and the higher fitness value determines the winning
programs. Tournament winners turn into individuals of the next population. \Biocode draws individual populations with replacement, and so drawn
individuals are copied in the next population. More fit programs have a higher chance to succeed in the tournaments,
so such members of the genetic algorithm  have a higher chance to survive into the subsequent generation.

\section{Applications to Real and Synthetic Biological Networks}

We evaluate the performance of our proposed framework \Biocode in
learning programs that generate graphs matching both real biological
networks and synthetic networks predefined attributes. We consider the
following \Biocode parameters in our experiments unless otherwise
noted. \Biocode programs in the optimization process first generation
start with randomly selected $10$ operations. Every generation comprises $100$ programs which are evaluated by
a single-objective or a multi-objective fitness functions as in
sections~\ref{sec:5.1} and~\ref{sec:5.2} respectively. \Biocode advances individual programs to next generations by
tournament. At the beginning of every generation, two-point crossover is used to
breed the individual population from the chosen individuals from
earlier generation. This crossover mechanism creates novel individual programs which can
be of different length. These individuals are then mutated with rate
$0.1$. After carrying out this optimization approach for $15$ generations,
the fittest program from the ultimate generation is chosen to be the
resulting representative \Biocode program. We compare other models
against this representative program. We have implemented \Biocode in Scala. \Biocode and datasets used in this paper are available at~\url{https://github.com/seferlab/biocode}.

\subsection{Learning scale-free graphs}\label{sec:5.1}

B-A model is motivated by generating graphs with scale-free node
degree distribution which is a fundamental property of many real-word
biological networks, and \Biocode is able to learn growth models that
generate scale-free distributions. Given the massive growth model space characterized by the operation
set and corresponding parameters, it is not clear whether effective exploration of that search space can
be achieved.

\Biocode uses a shape function to calculate the degree distributions
similarity. In this case, one can select the goodness of fit to a
scale-free distribution as an attribute. However, this method cannot be generalized to distributions other than
scale-free distributions. Generally, our goal is to produce graphs
matching an arbitrary degree distribution's shape. We model the shape of arbitrary distribution by defining the shape
$\psi_{shape}$ as the cumulative distribution of vertex degrees where
degrees of the vertices~(support of the distribution) is scaled to range $0$ and $1$. We can compare the degree distribution of different size graphs after such scaling.
Similarity measure for the shape attribute is defined as:
\begin{equation}\label{eq:4}
s_{shape}(\psi_{shape}^{i}, \psi_{shape}^{j}) =
\frac{1}{\norm{\psi_{shape}^{i} - \psi_{shape}^{j}}_{1} \,+\,\epsilon}
\end{equation}
where $\epsilon$, as a tiny positive constant, ensures that fitness is
well-defined when the compared shapes exactly match with each
other. The single parameter of B-A model is $i$ which is the number of existing nodes to
which a newly introduced node attaches. We retrieve the target node degree distribution shape for $i = 3, 4,
5, 6$ by producing graphs for such $i$ values and obtaining scale-free exponents maximum
likelihood estimates of $\alpha = 2.6, 2.7, 2.8, 2.9$ for each
$i$. Then, by utilizing $s_{shape}$ in Eq.~\ref{eq:3}, degree distribution shape difference between the estimated target
shape and the program generated graphs characterizes the program fitness.

By using this fitness function, \Biocode can learn multiple different
programs that generate scale-free graphs. One of the most effective
\Biocode programs that generate scale-free graph is shown in
Algorithm~\ref{alg:4}. We test the possibility of a scale-free degree distribution by using statistical
tests specificially designed for scale-free distribution as defined
in~\cite{Clauset2009}. Although \Biocode has not explicitly used the $\alpha$ parameter in
the fitness function, the mean $\alpha$  values of the graphs generated from the learned models
passing the scale-free test is $2.69$, that is fairly similar to the
target graphs $\alpha$.

\begin{algorithm}
\caption{Instance of Learned Scale-Free Model}
\begin{algorithmic}[1]
\State \Newnode
\State \Randomnode
\State \Attachtoinfluenced
\State \Clear r$2$
\State \Set$(1)$
\State \Randomedge
\State \Detachfrominfluenced
\State \Randomnode
\State \Createedge
\State \Influenceneighbors$(0.692)$
\end{algorithmic}
\label{alg:4}
\end{algorithm}

\begin{figure}[h!]
\begin{center}
  \includegraphics[width=0.5\textwidth]{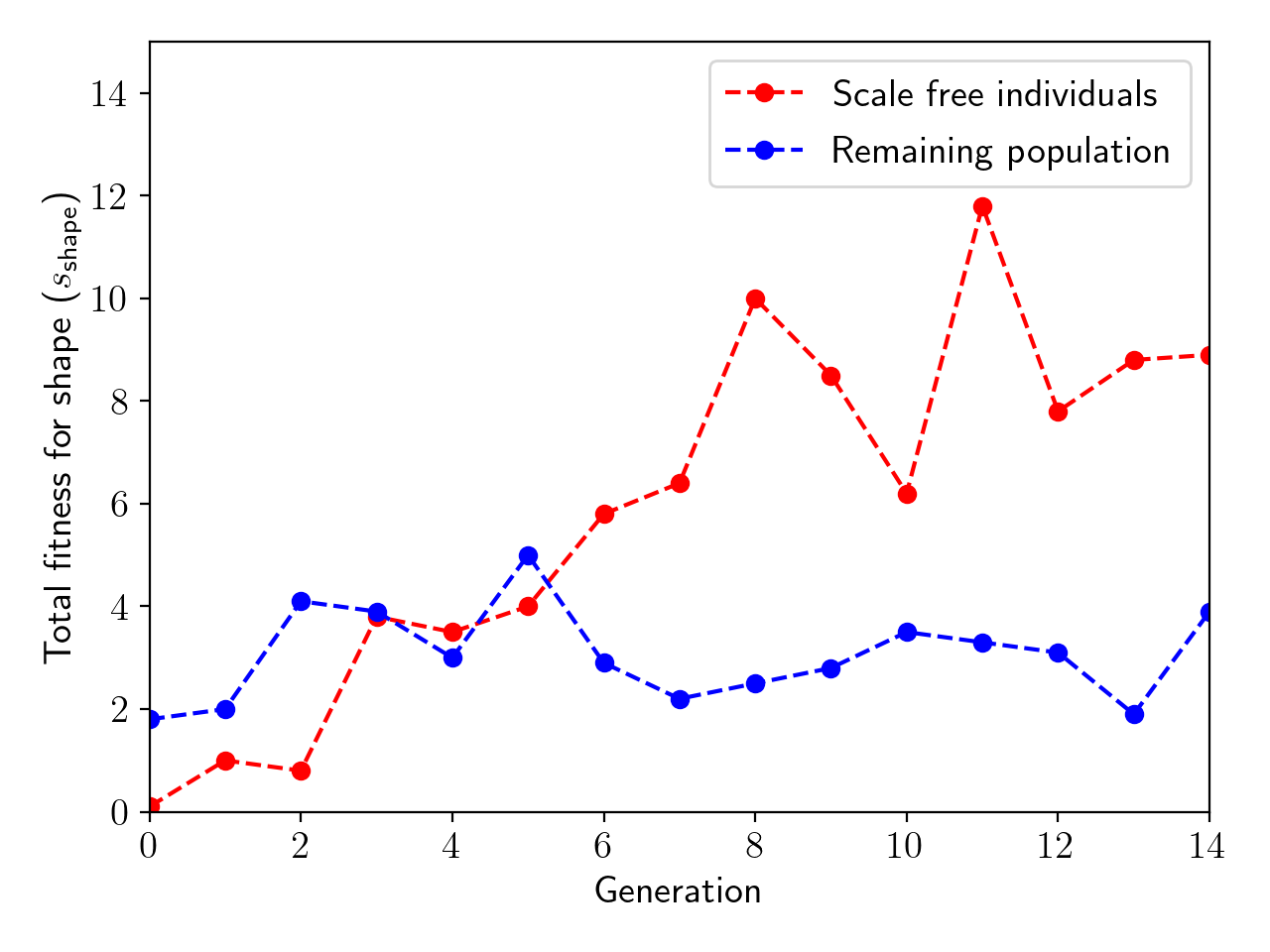}
\caption{Total fitness for shape function ($s_{shape}$) at each
  generation. The total fitness for individuals failing the scale-free test is
  shown in blue, whereas the total fitness for programs passing the
  scale-free test is plotted in red. Even after $5$th generation, scale-free individuals total fitness is
almost two times as big as the non-scale-free programs.}
\label{fig:2}
\end{center}
\end{figure}

Indeed, we claim that discovering a \Biocode program which generates
graphs with a scale-free degree distribution is reasonably easy for
the optimization process of \Biocode. One of the optimization trace while trying to fit a degree-distribution generated
from the B-A model with $i = 4$ is shown in Figure~\ref{fig:2}. According to Figure~\ref{fig:2}, \Biocode discovers scale-free models in the first generation even
before selection has started to make an impact on the population. Scale-free programs total fitness increases fast, and such total fitness is
considerably higher than the total fitness of the remaining individuals without scale-free distribution by generation $6$ as seen
in Figure~\ref{fig:2}. Our observations mainly indicate the following: 1- Scale-free model
discovery is not so challenging, 2- The possibility of scale-free
distribution in the graph appears to correlate quite well with the shape function.

\subsection{Performance on a biological collaboration network}\label{sec:5.2}

We test the performance of \Biocode over a co-authorship network of
genome-wide association studies (GWAS)~\cite{Bulik-Sullivan2012}. Particularly, we focus on such robust biological collaboration network
of “repeated co-authorship” where scientist pairs have an edge between
them if these scientists have published together more than one time. This biological collaboration network has high assortativity value of
$0.19$ showing that highly-collaborating scientists have an edge
with scientists that also collaborate profoundly. In this case, \Biocode concurrently optimizes for assortativity and
the shape distribution attributes by utilizing the multi objectivecprocedure discussed in section~\ref{sec:4.2}.

\begin{figure}[htbp]
 \begin{minipage}[b]{0.96\linewidth}
\centering
\includegraphics[width=\linewidth]{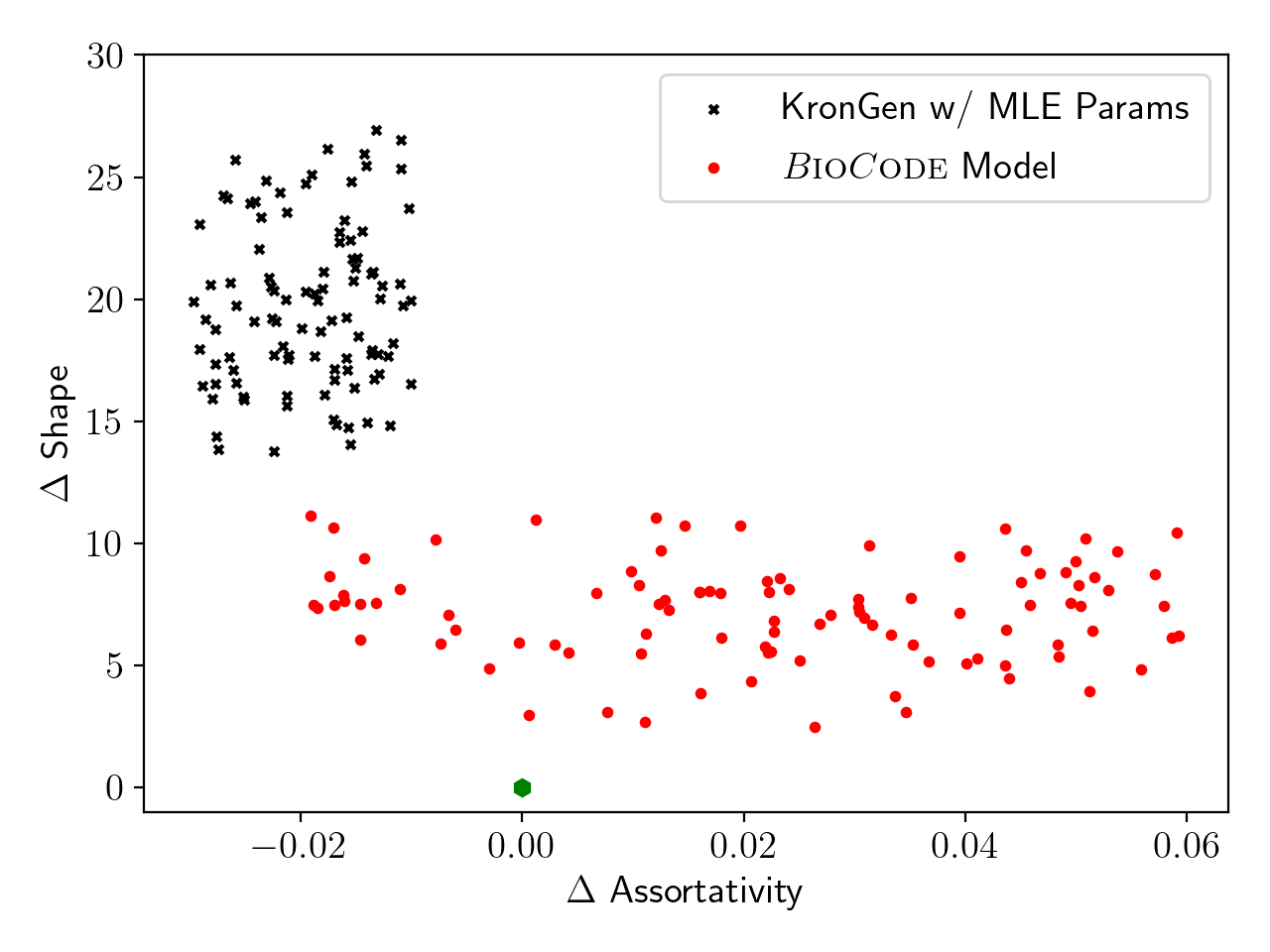}
\caption{GWAS biological collaboration target network.
Each point in the plot shows an individual produced graph from a
model. The $x$ axis shows the assortativity difference between the target
network and a graph. The $y$ axis shows the shape difference between the target
network and a graph. The green dot shows an exact match to the target network.}
\label{fig:3}
\end{minipage}
\end{figure}

We evaluate the performance by comparing the graphs produced by
\Biocode program to the graphs produced via Kronecker
model~\cite{Leskovec2010}. The Kronecker model recreates many real-world network attributes by its
recursive and fast procedure. We estimate Kronecker model parameters
by using the KronFit maximum likelihood method on the GWAS network. We compare real GWAS graph to the attributes of $100$ graphs produced
by each model. The learned \Biocode program outperforms the best-fit Kronecker model
in terms of better matching the degree distribution shape and the
assortativity of the true graph as in Figure~\ref{fig:3}. The mean shape difference of the \Biocode model is more similar to the
co-authorship network's shape than the shape for Kronecker model. The
average assortativity for Kronecker graphs is $0.165$ whereas the
average for \Biocode graphs is $0.206$. On the other hand, \Biocode generated graphs have a broader range of
assortativity scores (std. dev $0.0208$) than the Kronecker graphs scores
(std. dev $0.00629$).

\subsection{Performance on a protein interaction network}\label{sec:5.3}

\begin{figure}[htbp]
\begin{minipage}[b]{0.96\linewidth}
\centering
\includegraphics[width=\linewidth]{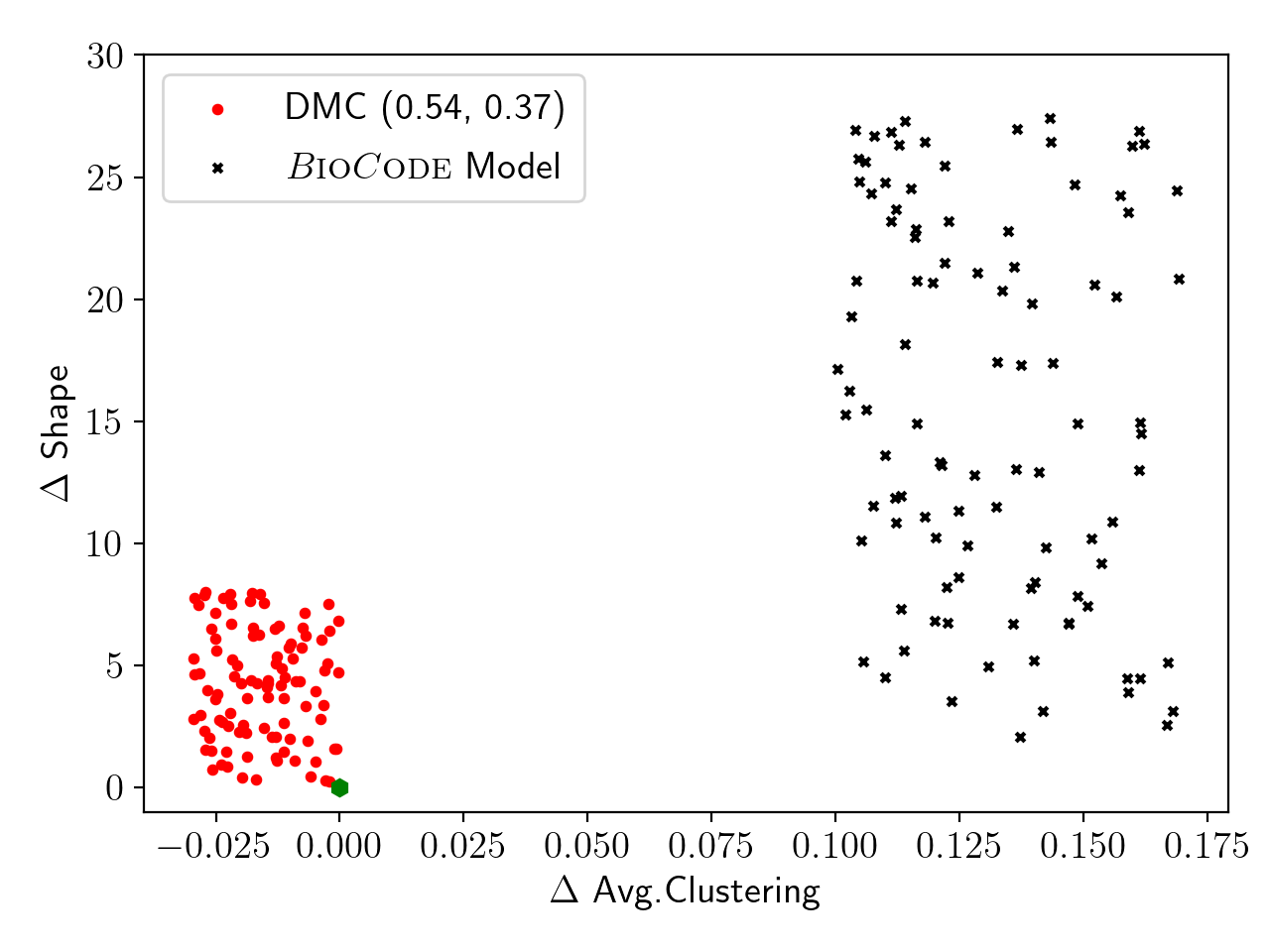}
\caption{Protein protein interaction target network. Each point in the plot shows an individual produced graph from a
model. The $x$ axis shows the average clustering coefficient difference between the target
network and a graph. The $y$ axis shows the shape difference between the target
network and a graph. The green dot shows an exact match to the target
interaction network.}
\label{fig:4}
\end{minipage}
\end{figure}

\Biocode is capable of learning a program that produces graphs similar
a recently compiled and high-quality yeast protein interaction
network~\cite{biogrid2021,Gibson2011}. \Biocode optimizes for both clustering coefficient and shape
distribution, which are biologically important in protein interaction
networks~\cite{Vazquez2011}. Our approach is similar the one described in Section~\ref{sec:5.2},
but DMC model is used for the baseline comparison instead of the
Kronecker model. The best parameters for DMC model are identified as $q_{con} = 0.37$
and $q_{mod} = 0.55$ via a grid search over the parameter space. The
parameters are chosen such that the graphs generated by \Biocode model
match the diameter, clustering coefficient, and match the number of
edges of the input protein interaction graph as close as possible. According to Figure~\ref{fig:4}, the graphs produced by \Biocode
program is considerably more similar to the real PPI network in terms
of target attributes than the ones generated by the DMC model. \Biocode program generates graphs with mean average clustering $0.091$
(standard deviation $0.006$) matching the true average clustering coefficient of
$0.099$ of the interaction network quite accurately. In contrast, the graphs generated by DMC model have mean average
clustering coefficient of $0.227$ (standard deviation $0.013$) that is truly far
away from the original interaction network's value. The shape
distribution results in Figure~\ref{fig:4} show the similar output. The average shape distribution distance of the random graphs produced
by the \Biocode program is $4.58$ (standard deviation $1.69$), where
the average distance between DMC graphs is $15.48$~(standard deviation
$6.29$). In addition to generating graphs better matching the target network,
\Biocode program show less variance and higher stability of parameters
in terms of those metrics.

\subsection{Performance on a gene regulatory network}\label{sec:5.4}

In the previous sections, \Biocode has outperformed the compared growth
model while optimizing models simultaneously for two network
attributes. Across different growth models, the particular network attributes were
selected for optimization since such attributes had already been
studied in the corresponding network class. However, \Biocode optimization is not limited to $2$ target
attributes. Here, we discuss the possibility of extending the learning
process to more than two attributes. We learn a \Biocode program over gene regulatory~(GR) network discussed
in~\cite{Lee2002, Leskovec2010} by simultaneously optimizing for all $3$
attributes such as average clustering coefficient, assortativity, and
shape. We evolve $150$ programs for $25$ generations over the gene
regulatory network. We compare the graphs produced by the optimized \Biocode program to the ones
produced by the Kronecker model as in Section~\ref{sec:5.2}.

\begin{figure}[h]
\begin{minipage}{.48\textwidth}
\centering
\includegraphics[scale=0.48]{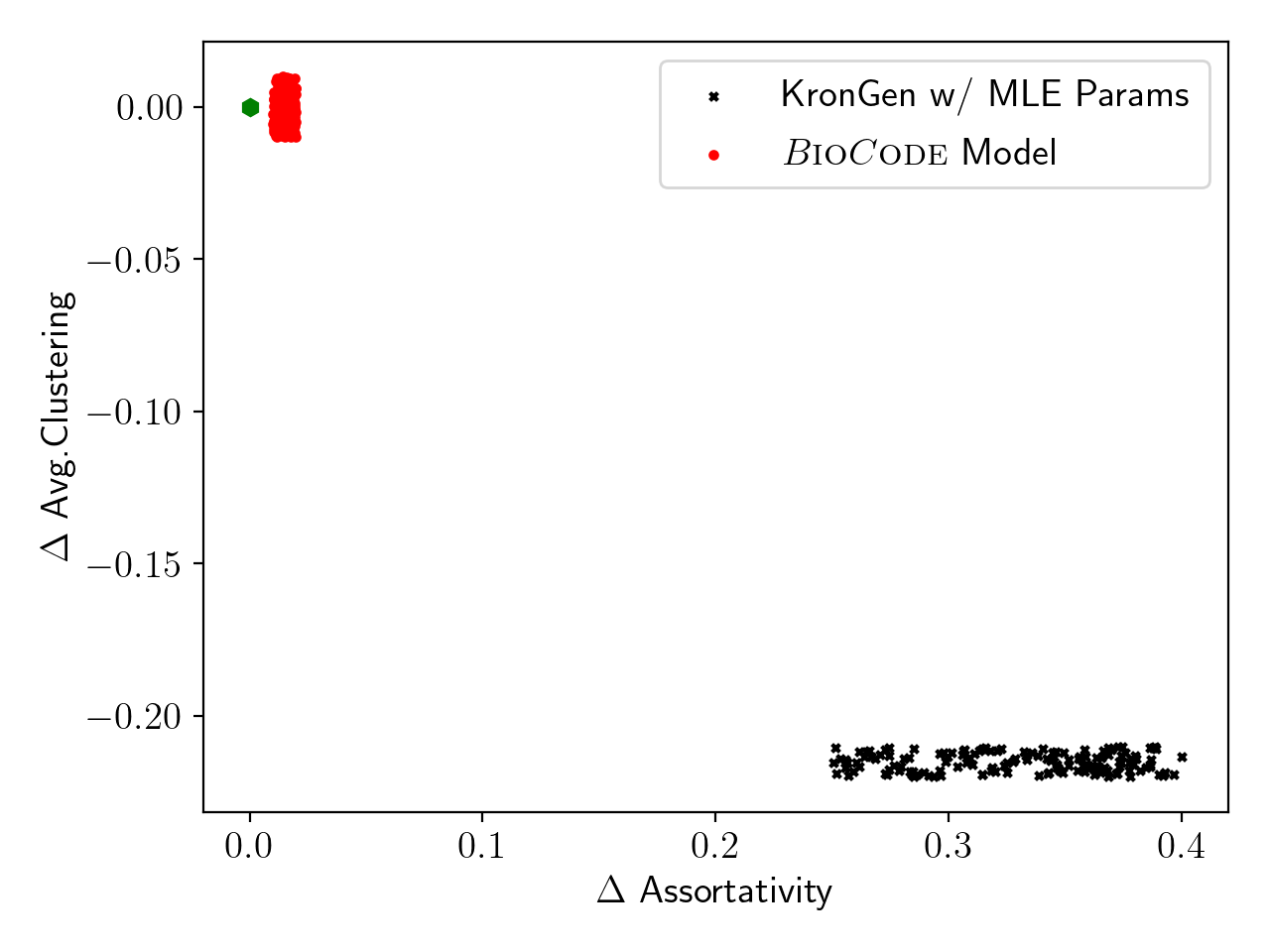}
\end{minipage}
\hfill
\begin{minipage}{.48\textwidth}
\centering
\includegraphics[scale=0.48]{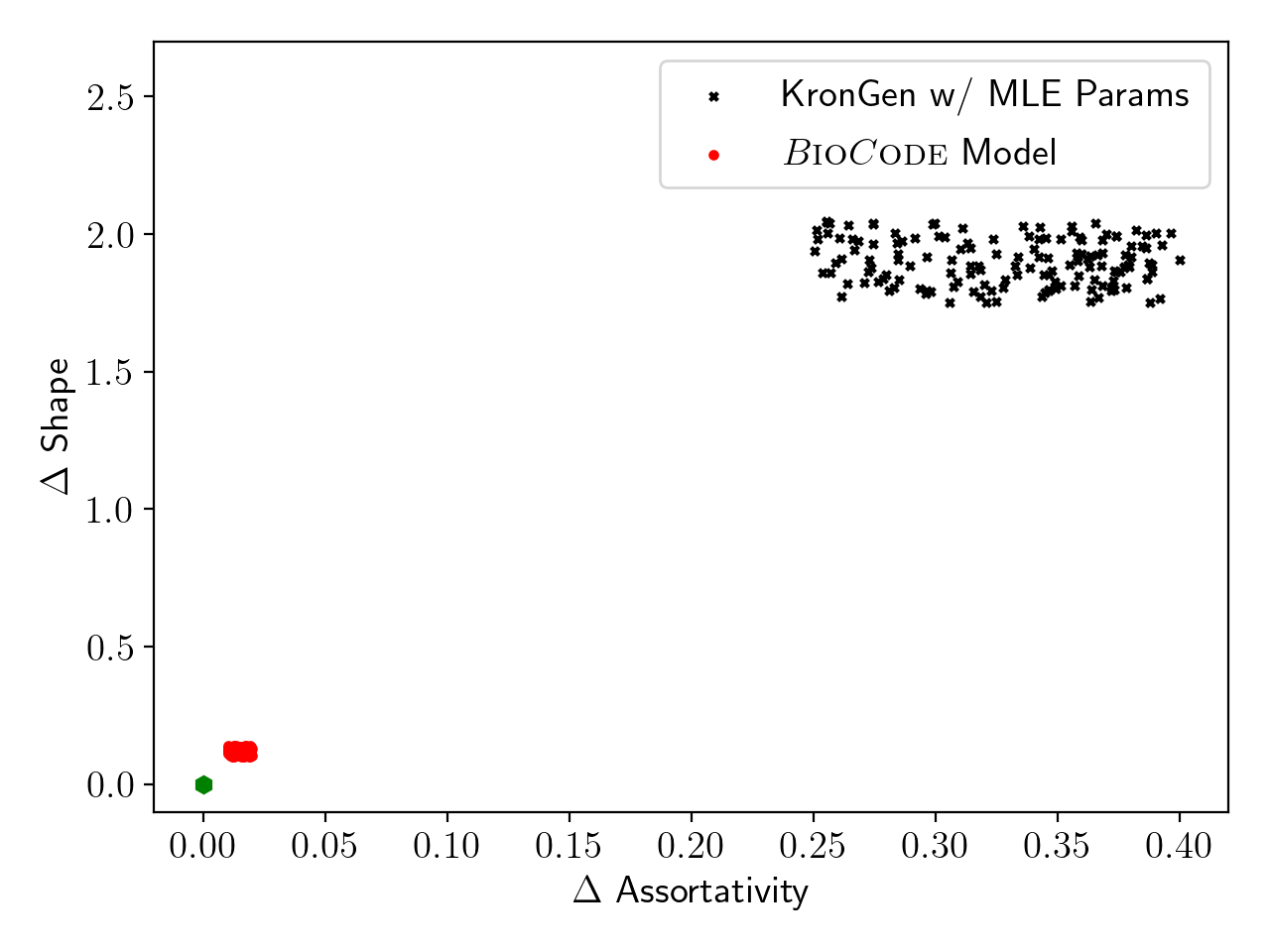}
\end{minipage}
\hfill
\begin{minipage}{.48\textwidth}
\centering
\includegraphics[scale=0.48]{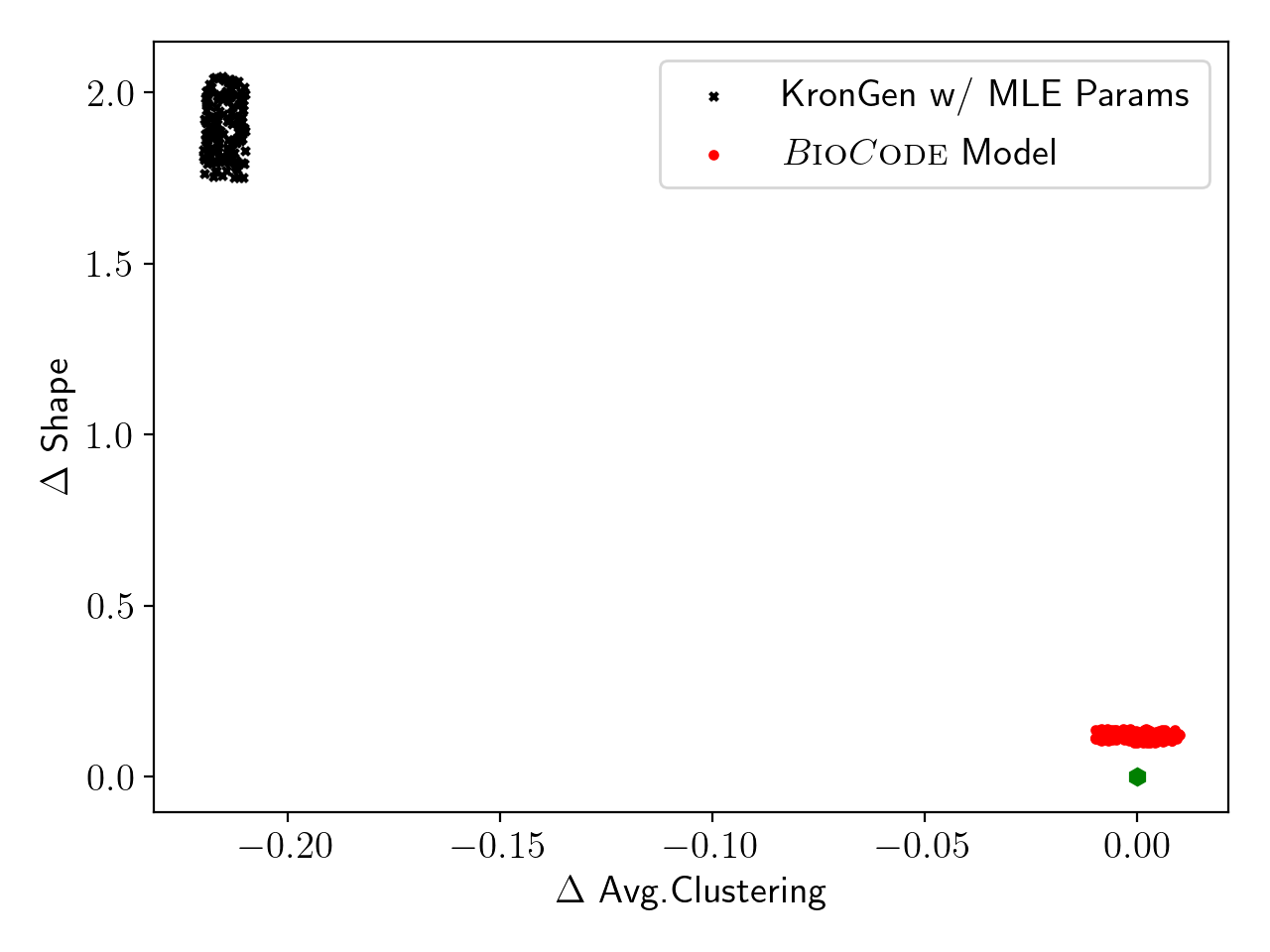}
\end{minipage}
\hfill
\caption{Gene regulatory target network. Each point in the plot
    shows an individual graph produced from a
    model. The difference between the coefficients of the gene regulatory
    network and a graph produced from the model are shown for all $3$
    network attribute pairs. The green dot defines the gene regulatory graph and shows the
    origin.}
  \label{fig:5}
\end{figure}

Figure~\ref{fig:5} includes $3$ plots which show the closeness of
\Biocode produced graphs to the gene regulatory network for all
attribute pairs. In reality, graphs generated by \Biocode outperforms the graphs produced by Kronecker model in terms of better
matching the gene regulatory graph. A single optimized \Biocode program was optimized with respect to all
$3$ attributes at a single time even though the plots in the figure
display $2$ dimensions at once. Graphs generated by \Biocode program have low
variance with respect to the target network topological properties, similar to the protein interaction network in Section~\ref{sec:5.3}.

\subsection{\Biocode generates random models}\label{sec:5.5}

The graphs generated by \Biocode program has higher diversity than the
ones generated by human-designed models. We use spectral distance between $100$ graphs produced by both the B-A model
and the \Biocode program to evaluate the diversity of graphs. The spectral distance is a plausible graph similarity metric
correlating highly with the graph edit distance~\cite{Wilson2008}. We calculated the spectral distances between graphs by using discretized
histogram of the normalized Laplacian eigenvalue distribution over
$100$ bins. In this case, the spectral distance becomes Euclidean
distance between such histograms.

\Biocode generates nondeterministic models as seen in
Table~\ref{tab:2}. Ensemble of graphs produced by \Biocode models have higher diversity
than the ensemble of graphs generated by the B-A model while matching
the target attributes better. We observe similarly higher diversity when we repeat this experiment
comparing the the graphs generated by\Biocode with the graphs produced
by DMC over yeast PPI network~\cite{biogrid2021}.

\begin{table*}[ht] 
\caption{Mean and standard deviation ($\mu \pm \sigma$) of spectral distance
between all graphs produced by \Biocode programs and by the B-A model.}
\begin{center}
\begin{tabular}{lllll}
\toprule
\parbox[b]{1.0cm}{\centering i} &\parbox[b]{2cm}{\centering 3} &
                                                      \parbox[b]{2cm}{\centering 4} &
 \parbox[b]{2cm}{\centering 5} & \parbox[b]{2cm}{\centering 6} \\
  \midrule
B-A & $0.0096 \pm 0.0068$ & $0.0044 \pm 0.0017$ & $0.0039 \pm 0.0016$
                                                   & $0.0036 \pm
                                                     0.0015$ \\
\Biocode  & $0.0151 \pm 0.0125$ & $0.0262 \pm 0.0216$ & $0.0298 \pm 0.0238$
                               & $0.0192 \pm 0.0162$ \\
\bottomrule
\end{tabular}
\end{center}
\label{tab:2}
\end{table*}

\section{Conclusions and Future Work}

We come up with \Biocode framework to represent network growth
dynamics as programs consisting of basic and expressive list of
operations. Such programs are general enough to closely approximate diverse set of
biological graph growth models. Besides, models with the desired attributes can be searched
effectively by combining efficient encoding of \Biocode with an
efficient genetic algorithm. Across gene regulatory, protein interaction, and biological
collaboration networks, the proposed optimization process is
reasonably fast: It takes less than $30$ minutes for $2$ objectives and less than $4$ hours
for $3$ objectives. In this setting, this optimization process can
generate \Biocode programs which compete strongly with the carefully-designed hand-coded network
growth models. Such hand-coded models are mainly introduced to match
graph attributes in related domains.

Graphs with scale-free degree distribution can be generated by \Biocode
for a number of attachment parameters $i$ such that these programs
pass stringent statistical tests to verify scale-free property.
In truth, \Biocode learning procedure discover scale-free programs
fast, and in the end, produces numerous different programs which generate graphs
passing the scale-freeness verification test. Moreover, \Biocode
generates graphs that are more varied than the graphs produced by the
B-A model. Overall, these results show the ubiquitousness of scale-free degree
distribution feature.

The introduced \Biocode framework generates unattributed graphs that
can be both directed and undirected, but it can be improved to
generate graphs with edge and node attributes as well. A number of enhancements are possible via expanding the instruction
set. As an example, one can add instructuons to diffuse node
attributes from one part of graph to remaining parts. More complicated
influence procedure can be incorporated to \Biocode as well. \Biocode machine needs to be enhanced to support these
instructions. For instance, \Biocode needs to add an edge memory
similar to the existing label memory to handle edge attributes. These enhacements are not so challenging, despite it is critical to design them carefully.

Lastly, even though individual instructions as part of \Biocode
programs can be interpreted quite easily, the growth dynamics of programs generated by the
learning process may not be so clear to a certain extent. As a future
work, we plan to focus on analyzing ensembles of optimized programs to
identify understandable growth mechanisms via identifying generally
appearing instruction motifs. For instance, by analyzing the programs similar to Algorithm~\ref{alg:4} in detail, we have identified several repeated
instruction patterns which can generate edges to current vertices
proportional to vertex degrees and can mimick scale-free graphs. Influence instructions, \Createedge, and \Newnode are commony
observed in these patterns. Finding instruction sets that are understandable as a unit can be
achieved by mining \Biocode programs for repeating instruction motifs.

% or


\bibliographystyle{IEEEtran} 
\bibliography{growcodebib}

\begin{IEEEbiography}[{\includegraphics[width=1in,height=1.25in,clip,keepaspectratio]{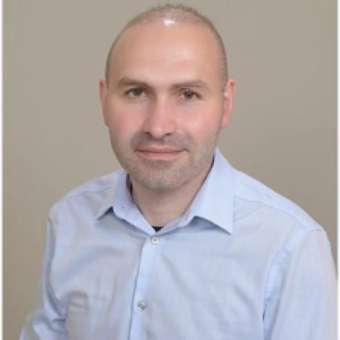}}]{Emre Sefer}
  obtained his B.Eng from Bogazici University,
  Department of Computer Engineering in 2008, M.S. in Computer Science
  from University of Maryland College Park in 2011, and Ph.D. in
  Computational Biology from Carnegie Mellon University in 2015. After
  completing his Ph.D. Dr. Sefer had a brief a post-doc at CMU Machine
  Learning Department with Ziv-Bar Joseph. He is
  currently an assistant professor in Computer Science Department,
  Ozyegin University. His academic research has
  focused on Bioinformatics, and Machine Learning applications on
  social and economic networks. He has published in number of journals
  and conferences during his PhD, receiving best research paper award
  at Recomb 2016 conference.
  %Since the end of 2015, Dr. Sefer has worked at Goldman Sachs and
  %J.P. Morgan as quantitative strategist and applied ML Researcher.
\end{IEEEbiography}


%\vfill


\end{document}